\documentclass{article}
\usepackage{dcase2019,amsmath,graphicx}
\usepackage{times,booktabs,tabularx,amsfonts}
\usepackage{etoolbox}
\usepackage[hyphens,spaces,obeyspaces]{url}
\usepackage[multiple]{footmisc}

\urldef{\dcasea}\url{http://www.cs.tut.fi/sgn/arg/dcase2016}
\urldef{\dcaseb}\url{http://www.cs.tut.fi/sgn/arg/dcase2017/challenge}
\title{Language Modelling for Sound Event Detection with Teacher Forcing and Scheduled Sampling}
\name{Konstantinos Drossos, Shayan Gharib, Paul Magron, and Tuomas Virtanen}
\address{Audio Research Group, Tampere University, Tampere, Finland\\
        \{firstname.lastname\}@tuni.fi}
\begin{document}
\ninept
\maketitle
\begin{sloppy}
\begin{abstract}
A sound event detection (SED) method typically takes as an input a sequence of audio frames and predicts the activities of sound events in each frame. In real-life recordings, the sound events exhibit some temporal structure: for instance, a ``car horn'' will likely be followed by a ``car passing by''. While this temporal structure is widely exploited in sequence prediction tasks (e.g., in machine translation), where language models (LM) are exploited, it is not satisfactorily modeled in SED. In this work we propose a method which allows a recurrent neural network (RNN) to learn an LM for the SED task. The method conditions the input of the RNN with the activities of classes at the previous time step. We evaluate our method using $F_{1}$ score and error rate ($ER$) over three different and publicly available datasets; the TUT-SED Synthetic 2016 and the TUT Sound Events 2016 and 2017 datasets. The obtained results show an increase of 9\% and 2\% at the $F_{1}$ (higher is better) and a decrease of $7\%$ and $2\%$ at $ER$ (lower is better) for the TUT Sound Events 2016 and 2017 datasets, respectively, when using our method. On the contrary, with our method there is a decrease of $4\%$ at $F_{1}$ score and an increase of $7\%$ at $ER$ for the TUT-SED Synthetic 2016 dataset. 
\end{abstract}
\begin{keywords}
sound event detection, language modelling, sequence modelling, teacher forcing, scheduled sampling
\end{keywords}
%
%
%
%
\section{Introduction}
Sound event detection (SED) consists in detecting the activity of classes (onset and offset times) in an audio signal, where the classes correspond to different sound events. (e.g., ``baby cry'', ``glass shatter''). This task finds applications in many areas related to machine listening, such as audio surveillance for smart industries and cities~\cite{Crocco:2016:ASS:2891449.2871183,Foggia2016}, smart meeting room devices for enhanced telecommunications~\cite{Butko2011,Busso2005}, or bio-diversity monitoring in natural environments~\cite{Furnas2015,Marques2013}. SED is a challenging research task since the sound events are of very diverse nature, which might be unknown a priori in real-life recordings. Besides, they often overlap in time, a problem termed as polyphonic SED. Significant advances in SED were made recently thanks to the advent of deep learning~\cite{Benetos2018}. The recurrent neural network (RNN) have proven particularly promising~\cite{Parascandolo2016,Hayashi2017} as they are able to model the temporal discriminant representations for sound events. More recently, these have been stacked with convolutional layers, resulting in convolutional recurrent neural networks (CRNN) which yield state-of-the-art results~\cite{cakir:2017:taslp,Adavanne2017}.

In real-life recordings, the various sound events likely temporal structures within and across events. For instance, a ``footsteps'' event might be repeated with pauses in between (intra-event structure). On the other hand, ``car horn'' is likely to follow or precede the ``car passing by'' sound event (inter-events structure). Although these temporal structures vary with the acoustic scene and the actual sound events classes, they exist and can be exploited in the SED task. Some previous studies focus on exploiting these temporal structures. For example, in~\cite{Hayashi2017}, the authors propose to use hidden Markov models (HMMs) to control the duration of each sound event predicted with a deep neural network (DNN). Although the results show some improvement with the usage of HMMs, the approach is a hybrid one and it requires a post processing step, which might be limited compared to an non-hybrid, DNN-based approach. In~\cite{Wang2017} and~\cite{gp:2018:iwaenc}, the connectionist temporal classification (CTC)~\cite{graves:2006:icml} loss function is used for SED: the output of the DNN is modified in order to be used with the CTC. Although the usage of CTC seems to be promising, CTC needs modification in order to be used for SED, it is a complicated criterion to employ, and it was developed to solve the problem where there is no frame-to-frame alignment between the input and output sequences~\cite{graves:2006:icml}. Thus, there might be the case that using a different method for SED language modelling could provide better results than CTC. Finally, in~\cite{gp:2018:iwaenc}, the authors also employ N-grams, which require pre and post processing stages, and use the class activities as extra input features. However, the latter approach did not perform better than a baseline which did not employ any language model.

In this paper we propose an RNN-based method for SED that exploit the temporal structures within and across events of audio scenes without the aforementioned drawbacks of the previous approaches. This method is based on established practices from other scientific disciplines that deal with sequential data (e.g., machine translation, natural language processing, speech recognition). It consists in using the output of the classifier as an extra input to the RNN in order to learn a model of the temporal structures of the output sequence (referred to as language model), a technique called \emph{teacher forcing}~\cite{williams:1989:tf}. Besides, this extra input of the RNN is chosen as a combination of the ground truth and predicted classes. This strategy, known as \emph{schedule sampling}~\cite{bengio:2015:nips}, consists in first using the ground truth activities and further replacing them by the predictions. This allows the RNN to learn a robust language model from clean labels, without introducing any mismatch between the training and inference processes.

The rest of the paper is organized as follows. In Section~\ref{sec:proposed-method} we present our method. Section~\ref{sec:evaluation} details the experimental protocol and Section~\ref{sec:results} presents the results. Section~\ref{sec:conclusions} concludes the paper. 
%
%
%
%
\section{Proposed method}\label{sec:proposed-method}
We propose a system that consists of a DNN acting as a feature extractor, an RNN that learns the temporal structures withing and across events (i.e. a language model), and a feed-forward neural network (FNN) acting as a classifier. Since we focus on designing an RNN that is able to learn a language model over the sound events, the RNN takes as inputs the outputs of both the DNN and the FNN. The code for our method can be found online\footnote{\url{https://github.com/dr-costas/SEDLM}}. 
\subsection{Baseline system}
The DNN takes as an input a time-frequency representation of an audio signal denoted $\mathbf{X}\in\mathbb{R}^{T\times F}_{\geq0}$, where $T$ and $F$ respectively denote the number of time frames and features. It outputs a latent representation:
\begin{equation}
    \mathbf{H} = \text{DNN}(\mathbf{X})\text{,}
\end{equation}
\noindent
where $\mathbf{H}\in\mathbb{R}^{T\times F'}$ is the learned representation with $F'$ features. Then, the RNN operates over the rows of $\mathbf{H}$ as
\begin{equation}\label{eq:rnn-layer}
    \mathbf{h}'_{t} = \text{RNN}(\mathbf{h}_{t}, \mathbf{h}'_{t-1})\text{,}
\end{equation}
\noindent
where $t=1,2,\ldots,T$, $\mathbf{h}'_{0} = \{0\}^{F''}$, $\mathbf{h}'_{t}\in[-1, 1]^{F''}$, and $F''$ is the amount of features that the RNN outputs at each time-step. Finally, the FNN takes $\mathbf{h}'_{t}$ as an input and outputs the prediction $\hat{\mathbf{y}}_{t}$ for the time-step $t$ as:
\begin{equation}
    \hat{\mathbf{y}}_{t} = \sigma(\text{FNN}(\mathbf{h}'_{t}))\text{,}
\end{equation}
\noindent
where $\sigma$ is the sigmoid function, and $\hat{\mathbf{y}}_{t}\in[0, 1]^{C}$ is the predicted activity of each of the $C$ classes.

The DNN, the RNN, and the FNN are simultaneously optimized by minimizing the loss $\mathcal{L}(\hat{\mathbf{Y}}, \mathbf{Y}) = \sum_t \mathcal{L}_{t}(\hat{\mathbf{y}}_{t}, \mathbf{y}_{t})$ with:
\begin{equation}
    \mathcal{L}_{t}(\hat{\mathbf{y}}_{t}, \mathbf{y}_{t}) = \sum\limits_{c=1}^{C}y_{t,c}\log(\hat{y}_{t,c}) + (1-y_{t,c})\log(1 - \hat{y}_{t,c})\text{,}
\end{equation}
\noindent
where $y_{t,c}$ and $\hat{y}_{t,c}$ are the ground truth and predicted activities, respectively, of the $c$-th class at the $t$-th time-step.
\subsection{Teacher forcing}
The modeling in Eq.~\eqref{eq:rnn-layer} shows that the RNN learns according to its input and its previous state~\cite{williams:1989:tf,bengio:2015:nips}. In order to allow the RNN to learn a language model over the output (i.e. the sound events), we propose to inform the RNN of the activities of the classes of the sound events at the time step $t-1$. That is, we condition the input to the RNN as:
\begin{equation}\label{eq:rnn-layer-tf}
    \mathbf{h}'_{t} = \text{RNN}(\mathbf{h}_{t}, \mathbf{h}'_{t-1}, \mathbf{y}'_{t-1})\text{,}
\end{equation}
\noindent
where $\mathbf{y}'_{t-1}$ is the vector with the activities of the classes of the sound events at time step $t-1$, and $\mathbf{y}'_{0} = \{0\}^{C}$. This technique is termed as teacher forcing~\cite{williams:1989:tf}, and is widely employed in sequence prediction/generation tasks where the output sequence has an inherent temporal model/structure (e.g., machine translation, image captioning, speech recognition)~\cite{bahdanau:2015:iclr,sutskever:2014:nips,vinyals:2015:cvpr}. By using this conditioning of the RNN, the RNN can learn a language model over the output tokens of the classifier~\cite{williams:1989:tf,bengio:2015:nips}. In SED, this results in letting the RNN learn a language model over the sound events, e.g., which sound events are more likely to happen together and/or in sequence, or how likely is a sound event to keep being active, given the previous activity of the sound events. Teacher forcing is different from what was proposed in~\cite{gp:2018:iwaenc}, as the latter approach conditioned the DNN (not the RNN) with the class activities: such an approach yielded poor results, intuitively explained by having $\mathbf{y}'_{t-1}$ dominated by the information in $\mathbf{X}$ through the sequence of the CNN blocks.
\subsection{Scheduled sampling}
The activity vector $\mathbf{y}'_{t-1}$ can be either the ground truth data (i.e., $\mathbf{y}_{t-1}$), or the predictions of the classifier (i.e., $\hat{\mathbf{y}}_{t-1}$). In the former case, the RNN is likely to start learning the desired language model from the first updates of the weights. At each time step $t$, the RNN will take as input the ground truth activities of the classes, thus being able to exploit this information from the very first weight updates. However, these ground truth values are not available at the inference stage: these would be replaced by the estimates $\hat{\mathbf{y}}_{t-1}$, which would create a mismatch between the training and testing processes. Besides, an RNN trained using only the true class activities is very likely to be sensitive to the prediction errors in $\hat{\mathbf{y}}_{t-1}$. Finally, we empirically observed that using $\mathbf{y}_{t-1}$ with the SED datasets, which are of relatively small size, results in a very poor generalization of the SED method.

A countermeasure to the above is to use the predictions $\hat{\mathbf{y}}_{t-1}$ as $\mathbf{y}'_{t-1}$, which allows the RNN to compensate for the prediction errors. However, during the first weight updates, the predicted $\hat{\mathbf{y}}_{t-1}$ is very noisy and any error created at a time step $t$ is propagated over time, which results in accumulating more errors down the line of the output sequence. This makes the training process very unstable and is likely to yield a poor SED performance.
\begin{figure}[!t]
    \centering
    \includegraphics[width=.71\columnwidth,trim={.0cm .0cm .0cm .0cm},clip]{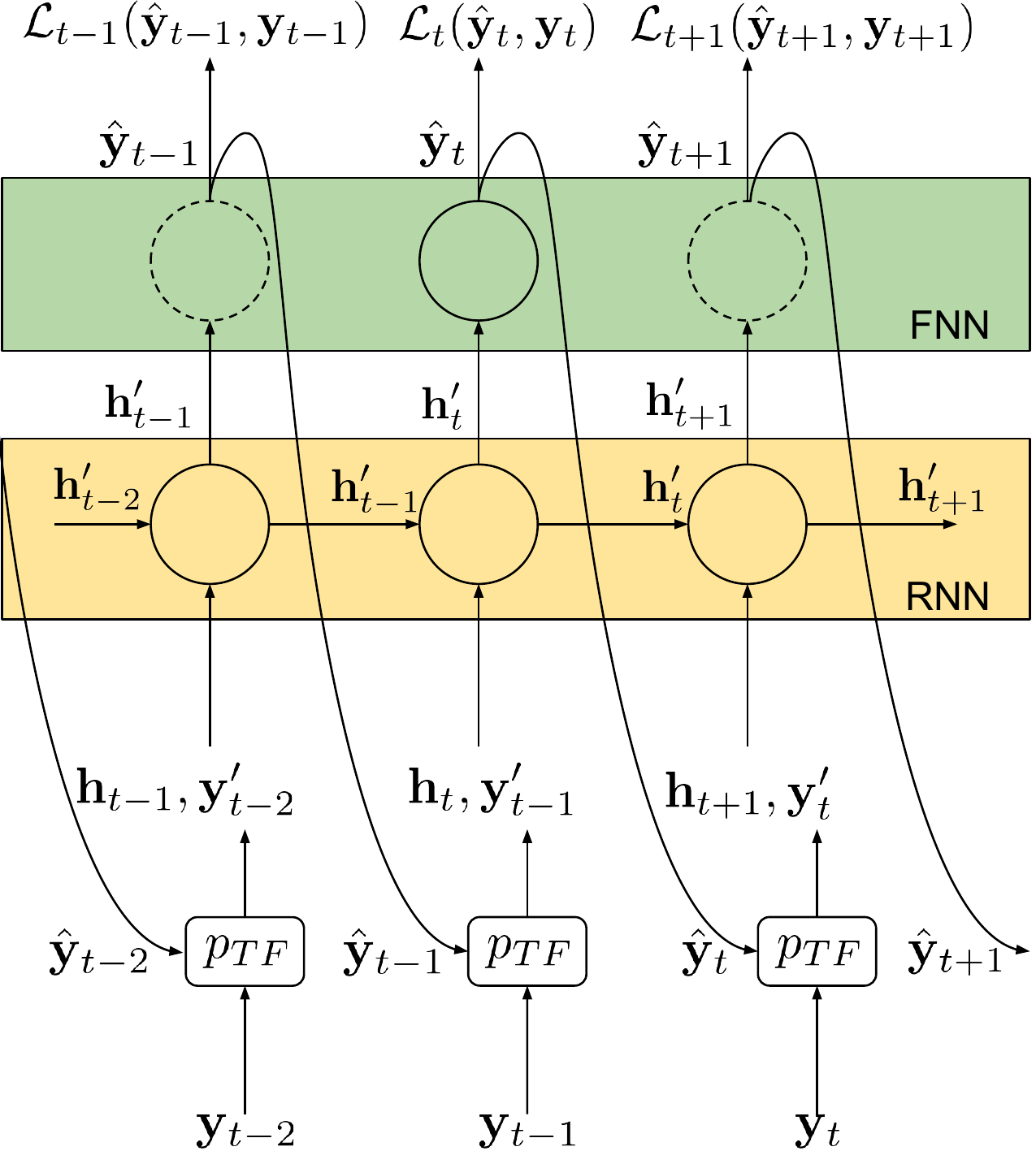}
    \caption{Proposed method of teacher forcing with scheduled sampling.}
    \label{fig:tf_method}
\end{figure}

To exploit the best of both approaches, we propose to use the scheduled sampling strategy~\cite{bengio:2015:nips}: the ground truth class activities are used during the initial epochs, and they are further gradually replaced by the predicted class activities. This gradual replacement is based on a probability $p_{\text{TF}}$ of picking $\mathbf{y}_{t-1}$ over $\hat{\mathbf{y}}_{t-1}$ as $\mathbf{y}'_{t-1}$ that decreases over epochs. Different functions can be used for the calculation of $p_{\text{TF}}$ (e.g., exponential, sigmoid, linear). Here, we employ a model of exponential decrease of $p_{\text{TF}}$:
\begin{equation}
    p_{\text{TF}} = \min(p_{\text{max}}, 1 - \min(1 - p_{\text{min}}, \frac{2}{1+e^{\beta}} - 1))\text{,}
\end{equation}
\noindent
where $\beta = -i\gamma N_{b}^{-1}$, $i$ is the index of the weight update (i.e., how many weight updates have been performed), $N_{b}$ is the amount of batches in one epoch, and $p_{\text{max}}$, $p_{\text{min}}$, and $\gamma$ are hyper-parameters to be tuned. $p_{\text{max}}$ and $p_{\text{min}}$ are the maximum and minimum probabilities of selecting $\hat{\mathbf{y}}_{t}$, and $\gamma$ controls the slope of the curve of $p_{TF}$ for a given $N_{b}$ and as $i$ increases. We use a minimum probability $p_{\min}$ because we experimentally observed that if we solely use $\mathbf{y}_{t-1}$ as $\mathbf{y}'_{t-1}$ even in the first initial weight updates, then the SED method overfits. The usage of $p_{\min}$ counters this fact. On the other hand, we use a maximum probability $p_{\max}$ in order to allow the usage of $\mathbf{y}_{t-1}$ as $\mathbf{y}'_{t-1}$ at the later stages of the learning process. We do this because the length of a sequence in SED is usually over 1000 time-steps and any error in $\hat{\mathbf{y}}_{t}$ is accumulated in this very long sequence, resulting in hampering the learning process. The usage of $p_{\max}$ offers a counter measure to this, by allowing the usage of ground truth values $\mathbf{y}_{t}$ instead of predicted and noisy values. This method is illustrated in Figure~\ref{fig:tf_method}. 
%
%
%
%
\section{Evaluation}\label{sec:evaluation}
We evaluate our method using the CRNN from~\cite{cakir:2017:taslp}, and we employ synthetic and real-life recordings datasets to illustrate the impact of the language model learned by our method.
\subsection{Data and feature extraction}
The synthetic dataset is the TUT-SED Synthetic 2016, used in~\cite{cakir:2017:taslp}, and consisting of 100 audio files which are synthetically created out of isolated sound events of 16 different classes. These classes are: \emph{alarms and sirens}, \emph{baby crying}, \emph{bird singing}, \emph{bus}, \emph{cat meowing}, \emph{crowd applause}, \emph{crowd cheering}, \emph{dog barking}, \emph{footsteps}, \emph{glass smash}, \emph{gun shot}, \emph{horse walk}, \emph{mixer}, \emph{motorcycle}, \emph{rain}, and \emph{thunder}. Each audio file contains a maximum of $N$ number of randomly selected target classes, where $N$ is sampled from the discrete uniform distribution $U(4,9)$, and the maximum number of simultaneously active (polyphony) sound events is 5. The audio files do not contain any background noise. The audio files amount to a total of 566 minutes of audio material, and according to the splits introduced by~\cite{cakir:2017:taslp}, roughly 60\% of the data are dedicated to training, 20\% to validation, and 20\% to testing split. More information about the dataset can be found online\footnote{\url{http://www.cs.tut.fi/sgn/arg/taslp2017-crnn-sed/tut-sed-synthetic-2016}}. 

We employ two real-life recording datasets, which were part of the Detection and Classification of Acoustic Scenes and Events (DCASE) challenge datasets for SED in real life audio task: the TUT Sound Events 2016 and the TUT Sound Events 2017~\cite{Mesaros2016_EUSIPCO}. The TUT Sound Events 2016 dataset contains sound events recorded in two environments: home (indoor), which contains 11 classes, and residential area (outdoor), which contains 7 classes. The classes for the home environment are: \emph{(object) rustling}, \emph{(object) snapping}, \emph{cupboard}, \emph{cutlery}, \emph{dishes}, \emph{drawer}, \emph{glass jingling}, \emph{object impact}, \emph{people walking}, \emph{washing dishes}, and \emph{water tap running}. The classes for the residential area environment are: \emph{(object) banging}, \emph{bird singing}, \emph{car passing by}, \emph{children shouting}, \emph{people speaking}, \emph{people walking}, and \emph{wind blowing}. The TUT Sound Events 2017 dataset contains recordings in a street environment and contains 6 different classes. These classes are: \emph{brakes squeaking}, \emph{car}, \emph{children}, \emph{large vehicle}, \emph{people speaking}, and \emph{people walking}. For both datasets, we use the cross-fold validation split proposed in the DCASE 2016 and 2017 challenges. More information about the classes, the cross-fold setting, and the recordings of the datasets can be found online\footnote{\dcasea}\footnote{\dcaseb}.

The synthetic dataset has randomly selected and placed sound events, therefore not exhibiting any underlying temporal structure of sound events. We thus expect the performance of our method to be similar to a method without language modeling on the synthetic dataset. Contrarily, the real-life recording datasets exhibit some underlying temporal structures in the sound events, therefore we expect our method to perform better than a method without language modelling on these datasets.

As input features $\mathbf{X}$ we use non-overlapping sequences of $T=1024$ feature vectors. These consist of $F=40$ log mel-bands, extracted using a short-time Fourier transform using a $22$ ms Hamming window, $50\%$ overlap and no zero padding. We normalize the extracted feature vectors from each dataset to have zero mean and unit variance, employing statistics calculated on the training split of each corresponding dataset.
\subsection{System and hyper-parameters}
As our DNN we use the three convolutional neural network (CNN) blocks from the system in~\cite{cakir:2017:taslp}, each consisting of a CNN, a batch normalization function, a max-pooling operation, a dropout function, and a rectified linear unit (ReLU). The kernels of the CNNs are square with a width of 5, a stride of 1, and a padding of 2 in both directions. There are 128 filters for each CNN. The kernel and the stride for the first max-pooling operation are $\{1, 5\}$, for the second $\{1, 4\}$, and for the third $\{1, 2\}$. These result in $F'=128$ for $\mathbf{H}$. All CNN blocks use a dropout of 25\% at their input, and the last CNN block also uses a dropout of 25\% at its output. As our RNN we use a gated recurrent unit (GRU) with $F''=128$ and our FNN is a single-layer feed-forward network with the output size defined according to the amount of classes in each dataset: $C=16$ for TUT-SED Synthetic 2016, $C=11$ and $C=7$ for the home and residential area scenes of the TUT Sound Events 2016, and $C=6$ for the TUT Sound Events 2017. To optimize the weights we employed the Adam optimizer~\cite{kingma:2015:adam} with default values. We employ a batch size of 8 and we stop the training when the loss for the validation data is not decreasing for 50 consecutive epochs. Finally, we set the hyper-parameters for $p_{\text{TF}}$ at $\gamma=12^{-1}$, $p_{\text{min}}=0.05$, and $p_{\text{max}}=0.9$. In Figure~\ref{fig:p_tf} is the value of $p_{\text{TF}}$ for consecutive weight updates of $N_{b}=44$ and for 100 epochs. 
\begin{figure}[!t]
    \centering
    \includegraphics[width=.85\columnwidth,trim={.1cm .1cm .1cm .1cm},clip]{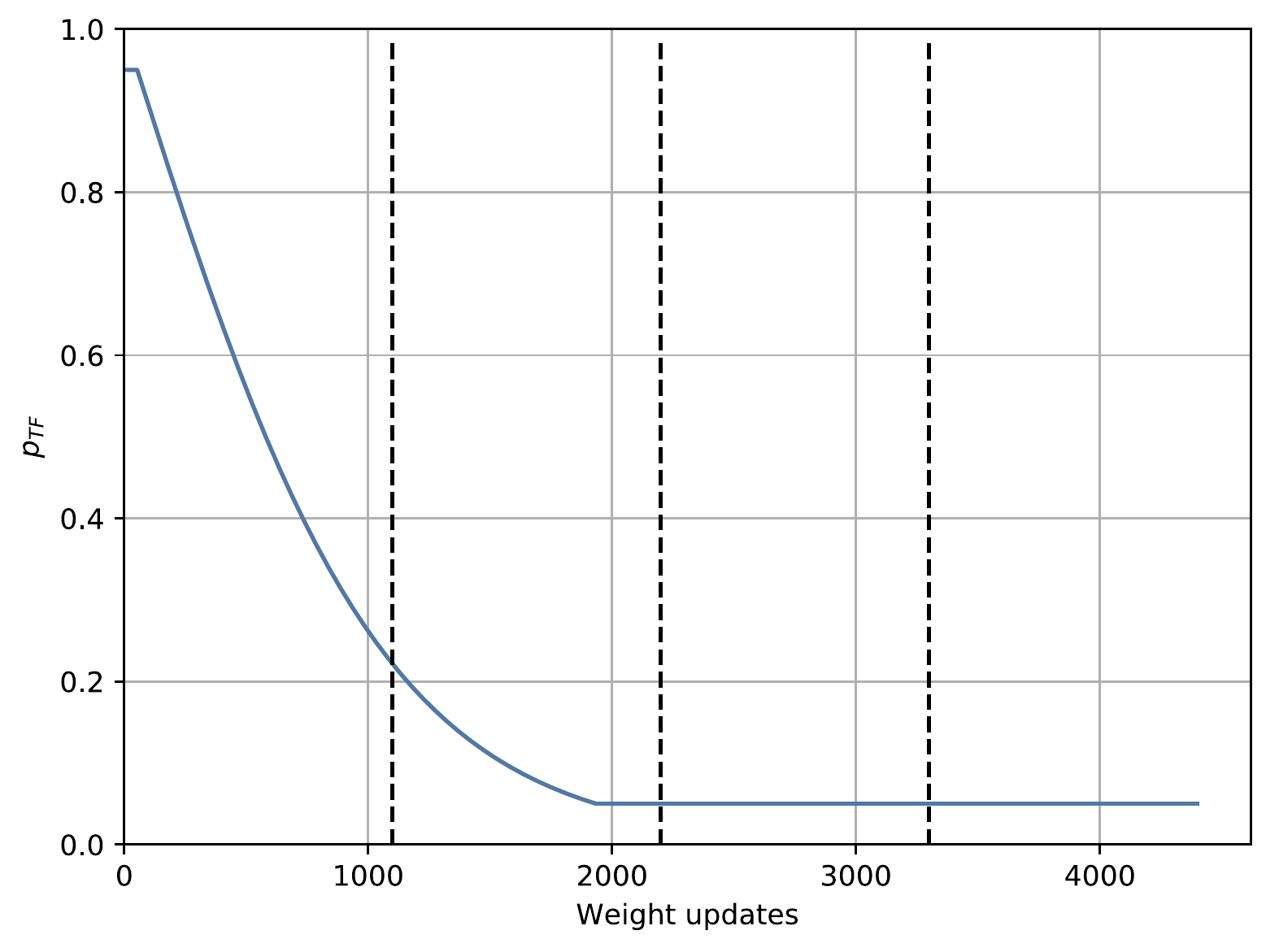}
    \caption{The value of $p_{\text{TF}}$ with consecutive weight updates with $p_{\min}=0.05$, $p_{\max}=0.95$, and $N_{b}=44$. The vertical dashed lines indicate steps of 25 epochs (i.e. 25, 50, 75 epochs).}
    \label{fig:p_tf}
\end{figure}

Empirically we observed that when using the TUT Sound Events 2017, there are some irregular spikes of relatively high gradients in different batches during training. To alleviate this issue, we clipped the $\ell_{2}$-norm of the gradient of all weights in each layer of our system to a value of 0.5. Additionally, we also observed that for the TUT Sound Events 2017 and TUT-SED Synthetic 2016 datasets, our method performed significantly better when we decreased the learning rate of the optimizer to $5e-4$. Therefore, we employed the above mentioned gradient clipping and modified learning rate for our method, when using the aforementioned datasets. Finally, for the TUT Sound Events 2016, we employed a binarized version of $\mathbf{y}'$ denoted $\mathbf{y}''$, such that $y_{t,c}''=1$ if $y_{t, c}'\geq 0.5$, and $y_{t,c}''=0$ otherwise.

All the above hyper-parameters were tuned using the cross validation set up for the TUT Sound Events 2016 and 2017 datasets provided by DCASE challenges, and the validation split provided in~\cite{cakir:2017:taslp} for the TUT-SED Synthetic 2016 dataset. 
\subsection{Metrics}
We measure the performance of our method using the frame based $F_{1}$ score and the error rate ($ER$), according to previous studies and the DCASE Challenge directions~\cite{cakir:2017:taslp,gp:2018:iwaenc}. For the real-life datasets, the $F_{1}$ and $ER$ are the averages among the provided folds (and among the different acoustic scenes for the 2016 dataset), while for the synthetic dataset the $F_{1}$ and $ER$ are obtained on the testing split. Finally, we repeat four times the training and testing process for all datasets, in order to obtain a mean and standard deviation (STD) for $F_{1}$ and $ER$.
\subsection{Baseline}
As a baseline we employ the system presented in~\cite{cakir:2017:taslp}, that does not exploit any language model. We do not apply any data augmentation technique during training and we use the hyper-parameters presented in the corresponding paper. This system is referred to as ``Baseline''.

When using our method with the TUT Sound Events 2017 and TUT-SED Synthetic 2016 datasets, we employ a modified learning rate for the optimizer and we clip the $\ell_{2}$-norm of the gradient for all weights. To obtain a thorough and fair assessment of the performance of our method, we utilize a second baseline for this dataset: we use again the system presented in~\cite{cakir:2017:taslp}, but we employ the above-mentioned gradient clipping and modified learning rate. We denote this modified baseline as ``modBaseline''. 

Finally, we compare our method to the best results presented in~\cite{gp:2018:iwaenc} which are obtained by employing N-grams as a post-processing to learn a language model. We report the results of this method on the TUT Sound Events 2016 datasets, as these are the only ones in the corresponding paper that are based on a publicly available dataset. It must be noted that in~\cite{gp:2018:iwaenc} was proposed the usage of $\mathbf{y}'_{t-1}$ as extra input features and the usage of CTC, but the results were inferior to the N-grams approach. Specifically, the per frame $F_{1}$ score was $0.02$ and $0.04$ lower and $ER$ was $0.02$ and $0.15$ higher with the usage of $\mathbf{y}'_{t-1}$ as an extra input feature and the usage of CTC, respectively, compared to the N-grams approach.
%
%
%
%
\section{Results \& discussion}\label{sec:results}
In Table~\ref{tab:results} are the obtained results for all the employed datasets. We remark that using the proposed language model improves the performance of SED in the real-life datasets. Specifically, for the TUT Sound Events 2016 dataset there is an improvement of $0.09$ in the $F_{1}$ score and $0.07$ for the $ER$. For the TUT Sound Events 2017, there is a $0.02$ improvement in $F_{1}$ and $0.02$ improvement in $ER$. These results clearly show that the employment of language modelling was beneficial for the SED method, when a real life datset was used. This is expected, since in a real life scenario the sound events exhibit temporal relationships. For example, ``people speaking'' and ``people walking'' or ``washing dishes'' and ``water tap running'' are likely to happen together or one after the other.

On the contrary, from Table~\ref{tab:results} we observe that there is a decrease in performance with our method on the synthetic data. Specifically, there is a $0.04$ (or $0.08$ when compared to modBaseline) decrease in $F_{1}$ and $0.07$ (or $0.12$ when compared to modBaseline) increase in $ER$. This clearly indicates that using a language model has a negative impact when the synthetic dataset is used. The sound events in the synthetic dataset do not exhibit any temporal relationships and, thus, the language model cannot provide any benefit to the SED method. We suggest that in such a scenario, the network focuses on learning a language model that does not exist in the data instead of solely trying to accurately predict the events on a frame-wise basis: this explains the drop in performance compared to the baseline method. Overall, this difference in performance between the two types of datasets strongly suggests that our method learns a language model over the activities of the sound events. 

Finally, our system significantly outperforms the previous method~\cite{gp:2018:iwaenc} on the TUT Sound Events 2016 dataset. This shows that learning a language model is more powerful than crafting it as a post-processing.
\begin{table}[!t]
\centering
\caption{Mean and STD (Mean/STD) of $F_{1}$ (higher is better) and $ER$ (lower is better). For the method~\cite{gp:2018:iwaenc} only the mean is available.}
\label{tab:results}\resizebox{.85\columnwidth}{!}{
\begin{tabular}{lcccc}
 & \multicolumn{1}{c}{\textbf{Baseline}} & \multicolumn{1}{c}{\textbf{modBaseline}} & \textbf{\cite{gp:2018:iwaenc}} & \multicolumn{1}{c}{\textbf{Proposed}} \\
 \hline
 & \multicolumn{4}{c}{TUT Sound Events 2016 dataset}\\
 \hline
 $\mathbf{F_{1}}$ & $0.28/0.01$ & \multicolumn{1}{c}{--} & $0.29$ & $0.37/0.02$ \\
 $\mathbf{ER}$ & $0.86/0.02$ & \multicolumn{1}{c}{--} & $0.94$ & $0.79/0.01$ \\
 \hline
 & \multicolumn{4}{c}{TUT Sound Events 2017 dataset}\\
 \hline
 $\mathbf{F_{1}}$ & $0.48/0.01$ & $0.49/0.01$ & \multicolumn{1}{c}{--} & $0.50/0.02$\\
 $\mathbf{ER}$ & $0.72/0.01$ & $0.70/0.01$ & \multicolumn{1}{c}{--} & $0.70/0.01$\\
 \hline
 & \multicolumn{4}{c}{TUT-SED Synthetic 2016 dataset}\\
 \hline
 $\mathbf{F_{1}}$ & $0.58/0.01$ & $0.62/0.01$ & \multicolumn{1}{c}{--} & $0.54/0.01$  \\
 $\mathbf{ER}$ & $0.54/0.01$ & $0.49/0.01$ & \multicolumn{1}{c}{--} & $0.61/0.02$ \\
\end{tabular}
}
\end{table}
\vspace{-6pt}
\section{Conclusions}\label{sec:conclusions}
\vspace{-3pt}
In this paper we presented a method for learning a language model for SED. Our method focuses on systems that utilize an RNN before the the last layer of the SED system, and consists of conditioning the RNN at a time step $t$ with the activities of sound events at the time step $t-1$. As activities for $t-1$ we select the ground truth early on the training process, and we gradually switch to the prediction of the classifier as the training proceeds over time. We evaluate our method with three different and publicly available datasets, two from real life recordings and one synthetic dataset. The obtained results indicate that with our method, the utilized SED system learned a language model over the activities of the sound events, which is beneficial when used on real life datasets. 

In future work, we will conduct a more in-depth analysis of the learned language model and of the SED performance per class. 
\bibliographystyle{IEEEtran}
\bibliography{refs}

\begin{thebibliography}{10}
\providecommand{\url}[1]{#1}
\def\UrlFont{\rmfamily}
\providecommand{\newblock}{\relax}
\providecommand{\bibinfo}[2]{#2}
\providecommand\BIBentrySTDinterwordspacing{\spaceskip=0pt\relax}
\providecommand\BIBentryALTinterwordstretchfactor{4}
\providecommand\BIBentryALTinterwordspacing{\spaceskip=\fontdimen2\font plus
\BIBentryALTinterwordstretchfactor\fontdimen3\font minus
  \fontdimen4\font\relax}
\providecommand\BIBforeignlanguage[2]{{%
\expandafter\ifx\csname l@#1\endcsname\relax
\typeout{** WARNING: IEEEtran.bst: No hyphenation pattern has been}%
\typeout{** loaded for the language `#1'. Using the pattern for}%
\typeout{** the default language instead.}%
\else
\language=\csname l@#1\endcsname
\fi
#2}}

\bibitem{Crocco:2016:ASS:2891449.2871183}
M.~Crocco, M.~Cristani, A.~Trucco, and V.~Murino, ``Audio surveillance: A
  systematic review,'' \emph{ACM Comput. Surv.}, vol.~48, no.~4, pp.
  52:1--52:46, Feb. 2016.

\bibitem{Foggia2016}
P.~{Foggia}, N.~{Petkov}, A.~{Saggese}, N.~{Strisciuglio}, and M.~{Vento},
  ``Audio surveillance of roads: A system for detecting anomalous sounds,''
  \emph{IEEE Transactions on Intelligent Transportation Systems}, vol.~17,
  no.~1, pp. 279--288, Jan 2016.

\bibitem{Butko2011}
T.~{Butko}, F.~G. {Pla}, C.~{Segura}, C.~{Nadeu}, and J.~{Hernando},
  ``Two-source acoustic event detection and localization: Online implementation
  in a smart-room,'' in \emph{Proc. European Signal Processing Conference}, Aug
  2011.

\bibitem{Busso2005}
C.~{Busso}, S.~{Hernanz}, {Chi-Wei Chu}, {Soon-il Kwon}, {Sung Lee}, P.~G.
  {Georgiou}, I.~{Cohen}, and S.~{Narayanan}, ``Smart room: participant and
  speaker localization and identification,'' in \emph{Proc. IEEE International
  Conference on Acoustics, Speech, and Signal Processing (ICASSP)}, March 2005.

\bibitem{Furnas2015}
B.~J. Furnas and R.~L. Callas, ``Using automated recorders and occupancy models
  to monitor common forest birds across a large geographic region,'' \emph{The
  Journal of Wildlife Management}, vol.~79, no.~2, pp. 325--337, 2015.

\bibitem{Marques2013}
T.~A. Marques, L.~Thomas, S.~W. Martin, D.~K. Mellinger, J.~A. Ward, D.~J.
  Moretti, D.~Harris, and P.~L. Tyack, ``Estimating animal population density
  using passive acoustics,'' \emph{Biological Reviews}, vol.~88, no.~2, pp.
  287--309, 2013.

\bibitem{Benetos2018}
E.~Benetos, D.~Stowell, and M.~D. Plumbley, \emph{Approaches to Complex Sound
  Scene Analysis}.\hskip 1em plus 0.5em minus 0.4em\relax Springer
  International Publishing, 2018, pp. 215--242.

\bibitem{Parascandolo2016}
G.~{Parascandolo}, H.~{Huttunen}, and T.~{Virtanen}, ``Recurrent neural
  networks for polyphonic sound event detection in real life recordings,'' in
  \emph{Proc. IEEE International Conference on Acoustics, Speech and Signal
  Processing (ICASSP)}, March 2016.

\bibitem{Hayashi2017}
T.~{Hayashi}, S.~{Watanabe}, T.~{Toda}, T.~{Hori}, J.~{Le Roux}, and
  K.~{Takeda}, ``Duration-controlled {LSTM} for polyphonic sound event
  detection,'' \emph{IEEE/ACM Transactions on Audio, Speech, and Language
  Processing}, vol.~25, no.~11, pp. 2059--2070, November 2017.

\bibitem{cakir:2017:taslp}
E.~{\c{C}akir}, G.~{Parascandolo}, T.~{Heittola}, H.~{Huttunen}, and
  T.~{Virtanen}, ``Convolutional recurrent neural networks for polyphonic sound
  event detection,'' \emph{{IEEE/ACM Transactions on Audio, Speech, and
  Language Processing}}, vol.~25, no.~6, pp. 1291--1303, June 2017.

\bibitem{Adavanne2017}
S.~{Adavanne}, P.~{Pertil\"{a}}, and T.~{Virtanen}, ``Sound event detection
  using spatial features and convolutional recurrent neural network,'' in
  \emph{Proc. IEEE International Conference on Acoustics, Speech and Signal
  Processing (ICASSP)}, March 2017.

\bibitem{Wang2017}
Y.~{Wang} and F.~{Metze}, ``A first attempt at polyphonic sound event detection
  using connectionist temporal classification,'' in \emph{Proc. IEEE
  International Conference on Acoustics, Speech and Signal Processing
  (ICASSP)}, March 2017.

\bibitem{gp:2018:iwaenc}
G.~{Huang}, T.~{Heittola}, and T.~{Virtanen}, ``Using sequential information in
  polyphonic sound event detection,'' in \emph{2018 16th International Workshop
  on Acoustic Signal Enhancement (IWAENC)}, Sep. 2018, pp. 291--295.

\bibitem{graves:2006:icml}
\BIBentryALTinterwordspacing
A.~Graves, S.~Fern\'{a}ndez, F.~Gomez, and J.~Schmidhuber, ``Connectionist
  temporal classification: Labelling unsegmented sequence data with recurrent
  neural networks,'' in \emph{Proceedings of the 23rd International Conference
  on Machine Learning}, ser. ICML '06.\hskip 1em plus 0.5em minus 0.4em\relax
  New York, NY, USA: ACM, 2006, pp. 369--376. [Online]. Available:
  \url{http://doi.acm.org/10.1145/1143844.1143891}
\BIBentrySTDinterwordspacing

\bibitem{williams:1989:tf}
R.~J. {Williams} and D.~{Zipser}, ``A learning algorithm for continually
  running fully recurrent neural networks,'' \emph{Neural Computation}, vol.~1,
  no.~2, pp. 270--280, June 1989.

\bibitem{bengio:2015:nips}
\BIBentryALTinterwordspacing
S.~Bengio, O.~Vinyals, N.~Jaitly, and N.~Shazeer, ``Scheduled sampling for
  sequence prediction with recurrent neural networks,'' in \emph{Proceedings of
  the 28th International Conference on Neural Information Processing Systems -
  Volume 1}, ser. NIPS'15.\hskip 1em plus 0.5em minus 0.4em\relax Cambridge,
  MA, USA: MIT Press, 2015, pp. 1171--1179. [Online]. Available:
  \url{http://dl.acm.org/citation.cfm?id=2969239.2969370}
\BIBentrySTDinterwordspacing

\bibitem{bahdanau:2015:iclr}
D.~Bahdanau, K.~Cho, and Y.~Bengio, ``Neural machine translation by jointly
  learning to align and translate,'' in \emph{International Conference on
  Learning Representations (ICLR)}, 2015.

\bibitem{sutskever:2014:nips}
\BIBentryALTinterwordspacing
I.~Sutskever, O.~Vinyals, and Q.~V. Le, ``Sequence to sequence learning with
  neural networks,'' in \emph{Advances in Neural Information Processing Systems
  27}, Z.~Ghahramani, M.~Welling, C.~Cortes, N.~D. Lawrence, and K.~Q.
  Weinberger, Eds.\hskip 1em plus 0.5em minus 0.4em\relax Curran Associates,
  Inc., 2014, pp. 3104--3112. [Online]. Available:
  \url{http://papers.nips.cc/paper/5346-sequence-to-sequence-learning-with-neural-networks.pdf}
\BIBentrySTDinterwordspacing

\bibitem{vinyals:2015:cvpr}
O.~{Vinyals}, A.~{Toshev}, S.~{Bengio}, and D.~{Erhan}, ``Show and tell: A
  neural image caption generator,'' in \emph{2015 IEEE Conference on Computer
  Vision and Pattern Recognition (CVPR)}, June 2015, pp. 3156--3164.

\bibitem{Mesaros2016_EUSIPCO}
A.~Mesaros, T.~Heittola, and T.~Virtanen, ``{TUT} database for acoustic scene
  classification and sound event detection,'' in \emph{24th European Signal
  Processing Conference 2016 (EUSIPCO 2016)}, Budapest, Hungary, 2016.

\bibitem{kingma:2015:adam}
D.~Kingma and J.~Ba, ``Adam: A method for stochastic optimization,'' in
  \emph{3rd International Conference for Learning Representations}, May 2015.

\end{thebibliography}
\end{sloppy}
\end{document}